\documentclass[runningheads]{svmult}

\usepackage{makeidx}   
\usepackage{graphicx}  
\usepackage{subeqnar}  
\usepackage{multicol}  
\usepackage{physprbb}  
\makeindex             

\newcommand{\oiiiaricher}{[O~{\sc iii}]$\lambda$4363}
\newcommand{\oiiinricher}{[O~{\sc iii}]$\lambda$5007}

\begin{document}
\title*{Planetary Nebulae as Probes of the Chemical Evolution
of Dynamically Hot Systems}
\toctitle{Planetary Nebulae as Probes of the Chemical Evolution of
Dynamically Hot Systems}
%
%
\titlerunning{Chemical Evolution of Dynamically Hot Systems}
%
\author{Michael G. Richer\inst{1}
\and Marshall L. McCall\inst{2}}
\authorrunning{M. G. Richer \& M. L. McCall}
%
%
\institute{Observatorio Astron\'omico Nacional,
Instituto de Astronom\'\i a, UNAM, P.O. Box 439027, San Diego, CA 92143
\and
Department of Physics and Astronomy, York University, 4700 Keele St.,
Toronto, Ontario, Canada  M3J 1P3}

\maketitle              

\begin{abstract}
The measurement of chemical abundances in planetary nebulae in
nearby galaxies is now relatively straightforward.  The challenge
is to use these chemical abundances to infer the chemical
evolution of their host galaxies.  At this point, our
understanding of chemical evolution based upon planetary nebulae
in galaxies without star formation is strongly coupled to our
understanding of the relationship between the chemical abundances
in stars and planetary nebulae in the Milky Way bulge. Supposing
that the same relation holds in all systems where star formation
ceased long ago, these systems follow a metallicity-luminosity
relation that is displaced to higher chemical abundances compared
to that found for dwarf irregular galaxies.  A more efficient
enrichment process appears to be required as part of the
explanation for this shift, in addition to the inevitable fading
of these galaxies.
\end{abstract}

\section{Introduction}

As a result of its evolution, a galaxy consumes gas to form
stars. Given a stellar initial mass function, the stars born in
any given generation will have a variety of lifetimes. The most
massive stars live only several $10^6$ years before exploding as
supernovae (SNe), their lifetimes being comparable to the time
scale for star formation.  When they explode as supernovae, these
stars return some fraction of their initial mass to the
interstellar medium (ISM).  Of this returned mass, some of it will
consist of elements synthesized during the star's evolution and so
the composition of the ISM will change over time as these
contributions accumulate.  One consequence is that the stars
forming at later times contain higher abundances of these
newly-synthesized elements.  Most stars contribute to this process
of enriching the ISM, though the time scales for the majority of
stars to contribute is of order several $10^9$ years.  The study
of the chemical evolution of galaxies attempts to characterize
this process of chemical enrichment as a function of galaxy type,
mass, and environment to better constrain theories of galaxy
formation and evolution.

Dynamically hot systems (DHSs) are stellar systems whose
kinematics are dominated by random motions, such as elliptical and
dwarf spheroidal galaxies and the bulges of spirals
\cite{benderetal92}.  Generally, DHSs have little gas and their
principal phase of star formation ceased several $10^9$ years ago.
Traditionally, the integrated spectrum of starlight has been used
to study the chemical evolution of these systems, though the
interpretation of integrated spectra is difficult due to the
interplay of the metallicities and ages of the stellar population
mix in any real galaxy.  Nonetheless, it has long been known that
these systems define a trend between the strength of their
absorption lines, e.g., the Mg$_2$ index, and their stellar
velocity dispersions, e.g., \cite{faber73,benderetal93}. This
trend is believed to arise due to winds that develop as a result
of the energy that supernovae inject into a galaxy's ISM while
star formation is still underway \cite{larson74}.

The lack of ongoing star formation in DHSs requires that another
probe besides H~{\sc ii} regions, the usual tool, be used to study
their chemical evolution. This is unfortunate, since, in H~{\sc
ii} regions, it is possible to measure the present abundances of
certain elements whose production appears easy to understand. It
is possible to study the chemical evolution of DHSs using stars or
planetary nebulae (PNe) as probes, e.g., \cite{tolstoyetal03},
provided that it is possible to observe elements whose abundances
are unchanged as a result of the probe's evolution. These probes
have the advantage over integrated spectra that their elemental
abundances may be obtained directly from their spectra. The
principal disadvantages of stars and PNe, as compared to H~{\sc
ii} regions, are determining their ages, and therefore the epoch
in a galaxy's evolution to which their chemical composition
corresponds, whether their chemical abundances have been altered
by stellar evolution, and at what enrichment level the process of
chemical evolution ceased in a given galaxy.  A wider range of
elemental abundances may be determined from the spectra of stars
than from those of PNe, but stellar spectra are more difficult to
measure to the required precision unless the stars are much
brighter than the galaxy background or unless the latter is faint.
In the central parts of galaxies, it is easier to obtain spectra
of PNe, which stimulates interest in their use for studying the
chemical evolution of DHSs.

Recently, \cite{mccallricher03} reviewed this topic.  That review
focussed more upon some of the theoretical underpinnings of
studying the chemical evolution of DHSs. Here we shall focus more
on issues related to observations.

\section{Spectroscopy of extragalactic planetary nebulae}

To date, published data have come from 4m-class telescopes.
8m-class telescopes should allow the derivation of oxygen
abundances from measured electron temperatures for galaxies out to
at least 5\,Mpc, and further if high spectral and spatial
resolution are used for measuring \oiiiaricher.

A variety of observational difficulties must be confronted when
obtaining spectroscopy of extragalactic PNe. The first arises from
differential atmospheric refraction over the wide wavelength
interval that must be observed
\cite{jacobyciardullo99,walshetal99}. Multi-object spectrographs
are an attractive option to increase the efficiency of the
observations by observing many PNe in a galaxy at once, but many
of these instruments lack atmospheric dispersion compensators and,
unlike in the case of long slit spectroscopy, it is not possible
to orient multi-object spectrographs so that the slitlets align
with the parallactic angle.  The second difficulty is subtracting
the light from the background galaxy, which is often much brighter
than the sky. The stellar absorption lines coincide with the
nebular Balmer lines and can vary in intensity on small spatial
scales. If there is an underlying galaxy disk, the diffuse
interstellar medium and its small-scale variation further
complicate the background subtraction by emitting a nebular
spectrum whose line intensities can be dramatically different from
those in the PNe under study.  Emission lines from such gas
preferencially comes from low ionization states, e.g., H~{\sc i},
[O~{\sc ii}], [N~{\sc ii}], and [S~{\sc ii}], but [O~{\sc iii}] is
also detected, though He~{\sc i} frequently is not
\cite{greenawaltetal97}. Emission from the night sky near He~{\sc
i}$\lambda$5876 and H$\alpha$-[N~{\sc ii}] is detrimental too,
though it is spatially uniform, in constrast to the galaxy
background. \cite{rothetal04} provide an instructive discussion of
this issue and argue convincingly of the advantages of 3D
spectroscopy, especially the advantage of having a 2D spatial map
of the background emission to better subtract this emission. A
third difficulty is the weakness of the \oiiiaricher\ line, which
is about 100 times fainter than \oiiinricher\ \cite{richeretal99},
and its proximity to the Hg~{\sc i}$\lambda$4358 line at
observatories whose night sky emission is contaminated by street
lights. \oiiiaricher\ must be observed if the electron temperature
is to be measured and used to obtain a reliable estimate of the
oxygen abundance.

It is difficult to characterize the absolute precision of the line
intensities measured to date in extragalactic PNe and the chemical
abundances derived from them. The most extensive comparison may be
made using the PNe studied in M32 by
\cite{richeretal99,richermccall04}, who used the same equipment
and refinements of the same observational technique. For lines
stronger than H$\beta$, the uncertainty appears to be about $\pm
25\%$.  For lines whose strength is 50\% that of H$\beta$, the
uncertainty appears to be about a factor of two. However, in all
cases, the abundances derived in both studies agree within
uncertainties. These \lq\lq internal" uncertainties can be as low
as $\pm 15\%$ and $\pm 0.1$\,dex for He and O abundances,
respectively, but may underestimate the true uncertainties if they
do not include all systematic uncertainties.

There are few observations of a given object by different groups.
In M31, PN29 is common to three studies.  The values found for the
oxygen abundance, the Ne/O ratio, and the N/O ratio all span a
range of about 0.3\,dex
\cite{richeretal99,jacobyciardullo99,rothetal04}. The helium
abundance differs by almost 40\% between two of the studies
\cite{jacobyciardullo99,rothetal04}.  PN23 has been observed twice
and the oxygen abundances and Ne/O ratios agree within
uncertainties \cite{richeretal99,jacobyciardullo99}. PN27 has also
been observed twice \cite{jacobyciardullo99,rothetal04}, with the
resulting oxygen abundances in excellent agreement, but the helium
abundances differing by 35\%. Although it is an indirect
indicator, there are several detections of He~{\sc i}$\lambda$5876
in PNe in M31 that are either anomalously strong or weak, perhaps
as a result of significant background contamination
\cite{richeretal99,jacobyciardullo99}.  In M32, the line
intensities for PN1 are in generally good agreement among the
available measurements
\cite{jenneretal79,richeretal99,richermccall04}. Considering all
of the data available, the case of PN29 in M31 appears unusual.
Were the oxygen abundances generally as uncertain as those
measurements imply, dispersions in abundances within a given
galaxy should be at least 0.3\,dex. In M32, NGC 185, and NGC 205,
the dispersions in the oxygen abundances for PNe with measured
electron temperatures are 0.19\,dex, 0.27\,dex, and 0.37\,dex,
respectively \cite{richeretal99,richermccall04}. A dispersion as
low as that found for M32 (13 PNe) implies that the typical
uncertainty in oxygen abundances should be less than 0.2\,dex,
which is comparable to the uncertainties quoted for individual
abundances (when quoted).

It appears that background subtraction may be the dominant source
of uncertainty in the line intensities at the moment
\cite{rothetal04}.  Currently, there are too few measurements to
determine whether the differences among studies are systematic.
Helium abundances appear to be uncertain by up to about 40\%,
while oxygen abundances may be uncertain by up to 60\%. Abundance
ratios of neon and nitrogen relative to oxygen may be less
uncertain, but one can anticipate difficulties with the background
subtraction for nitrogen lines (and eventually those of sulphur).
In principle, observations with higher spectral resolution should
help, since the PNe should have velocities that scatter about the
velocity of the diffuse ionized gas in the ISM, allowing
separation of their components.

What matters, however, for studies of the chemical evolution of
galaxies are the mean abundances and their dispersions for a
population of PNe. Provided that the oxygen abundances for
individual PNe in current studies do not deviate systematically
from their true values, the mean value for a sample of PNe should
be robust. On the other hand, the dispersions in abundances may be
more seriously affected if the true uncertainties are severely
underestimated.

\section{Interpreting the chemical abundances in bright planetary nebulae}

The first issue that has an important bearing upon the
interpretation of the abundances measured in PNe is the object
selection process.  To derive the most secure oxygen abundances,
the electron temperature must be measured, requiring that
\oiiiaricher\ be detected.  Since this line is about 100 times
fainter than \oiiinricher, the sample will normally be restricted
to PNe that are bright in \oiiinricher.  \lq\lq Bright PNe" is
adopted to denote PNe bright in \oiiinricher, usually within
2\,mag of the peak of the PN luminosity function.  To use this
population of PNe to infer the chemical evolution of their host
galaxies, it is necessary to understand the biases intrinsic to
this population of PNe.

\begin{figure}[t]
\begin{center}
\includegraphics[bb=160 270 560 720,width=5cm]{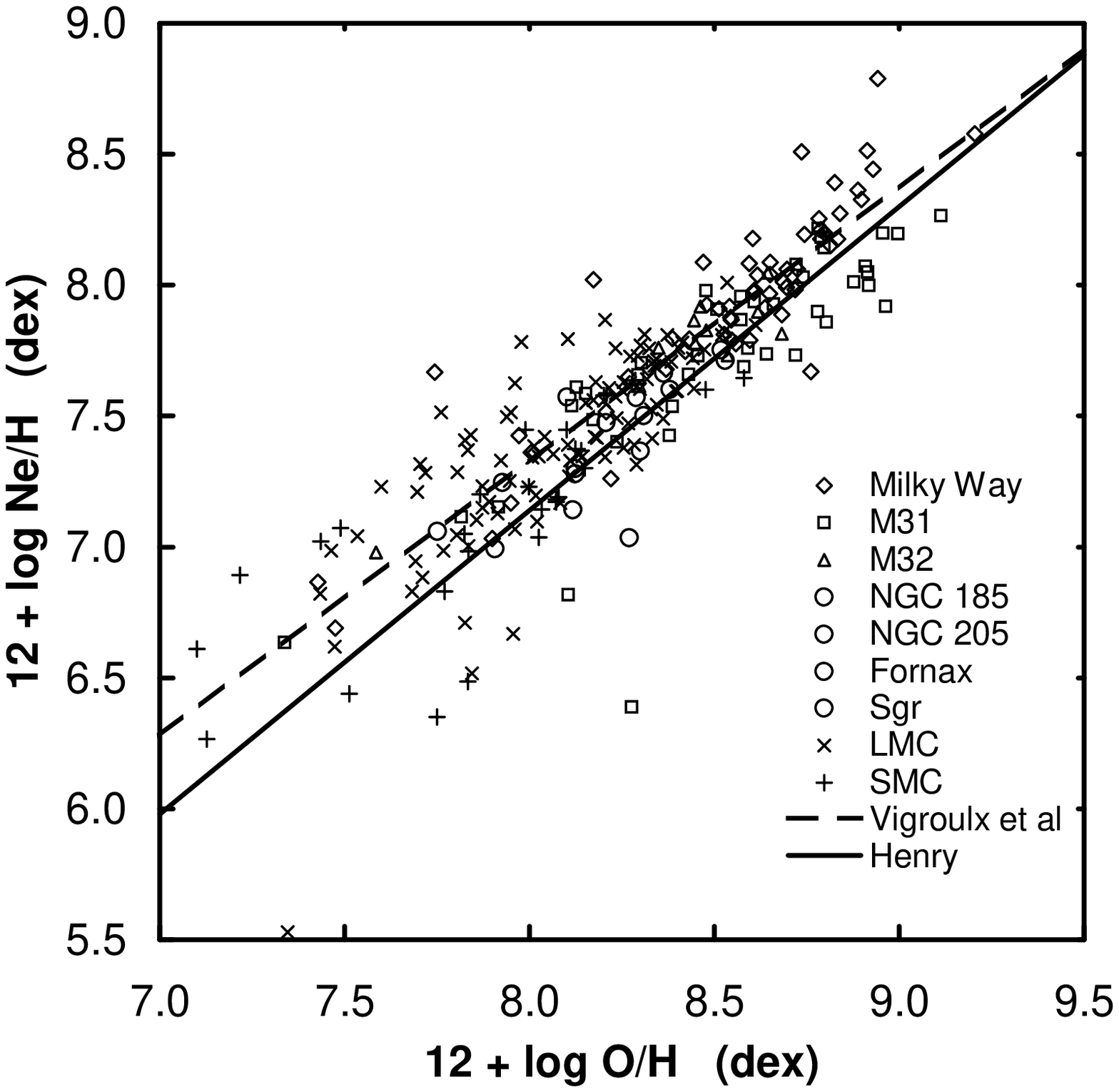}
\includegraphics[bb=70 270 470 720,width=5cm]{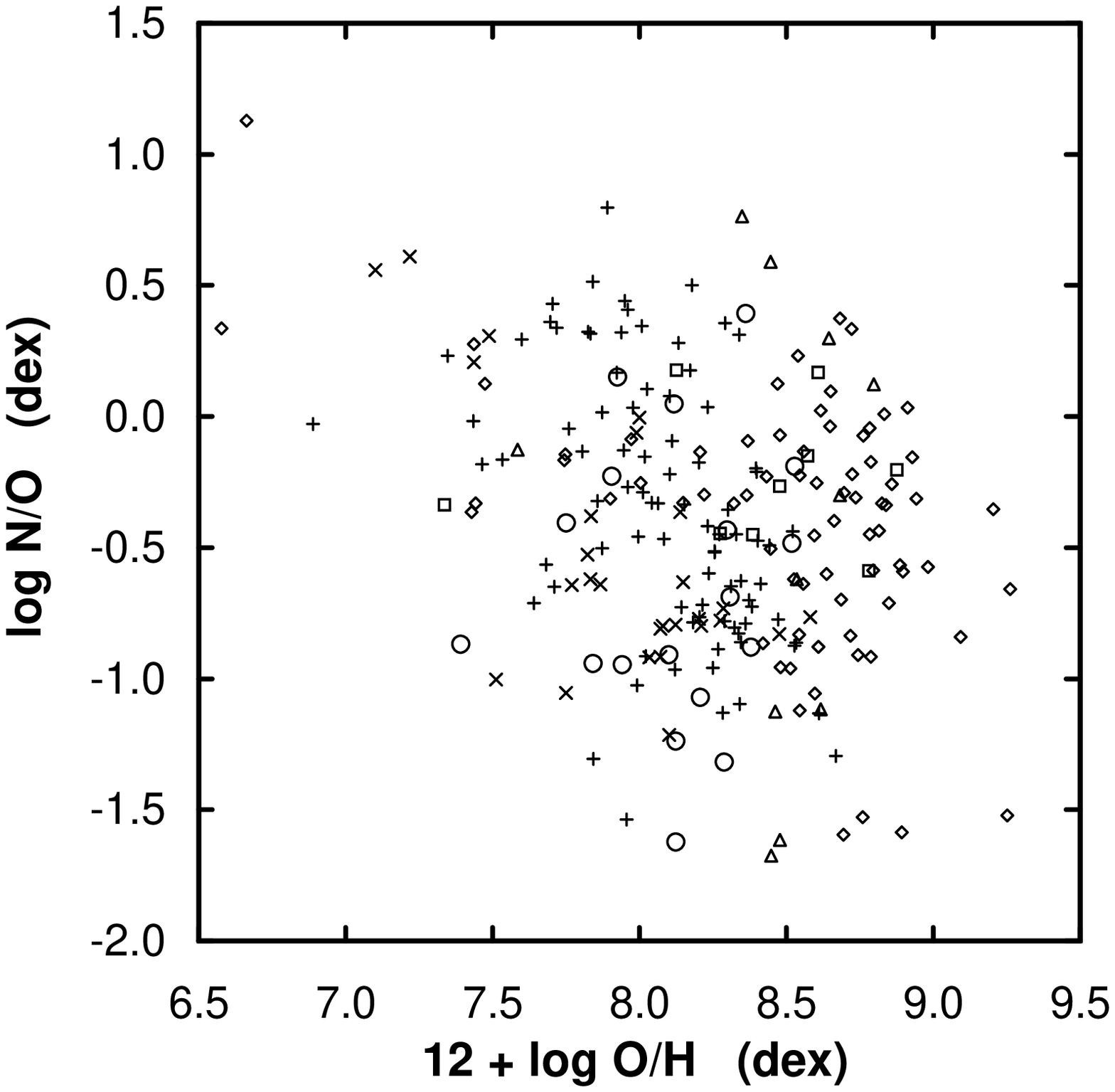}
\end{center}
\caption[]{Left: The oxygen and neon abundances observed in bright
PNe follow the same relation as found for the ISM in star-forming
galaxies \cite{vigroulxetal87,henry89}.  The simplest
interpretation is that the progenitors of bright PNe do not
significantly modify either of these abundances.  Right: The N/O
ratios for PNe in all galaxies scatter over the same range of
values, indicating that nucleosynthesis and mixing processes in
the progenitor stars of bright PNe have approximately the same
effect in all galaxies. The symbols have the same meaning in both
figures.  The data are adopted from
\cite{stasinskaetal98,jacobyciardullo99,rothetal04}.}
\label{fig_neo}
\end{figure}

Second, planetary nebulae are the descendants of evolved stars. It
is necessary to know how the nucleosynthesis associated with the
evolution of the progenitor star has modified its original
chemical composition, for only elements whose abundance survives
this evolution unchanged are useful for studying the chemical
evolution of the host galaxies. It is well-established that the
stellar progenitors of PNe modify their original abundances of He,
C, N, and s-process elements, e.g.,
\cite{marigoetal96,forestinicharbonnel97}.  Fig. \ref{fig_neo}
implies that the processes of nucleosynthesis and dredge-up in PN
progenitors appear to be approximately equally efficient in all
galaxies studied so far, since the N/O ratios in their bright PNe
span the same range of values. There is still considerable debate
regarding whether O and Ne are modified. Models indicate that PN
progenitors may modify their initial abundance of O and Ne
slightly, though the magnitude and even the sign of the effect is
unclear \cite{marigoetal96,forestinicharbonnel97}.
Observationally, the abundances of O and Ne in bright PNe
correlate in the same way as in giant extragalactic H~{\sc ii}
regions (Fig. \ref{fig_neo})
\cite{vigroulxetal87,henry89,stasinskaetal98}. Type II supernovae
are believed to be primarily responsible for setting this
abundance ratio in the ISM, so observing the same correlation in
bright PNe and H~{\sc ii} regions implies that the progenitors of
bright PNe probably do not significantly modify either their O or
Ne abundances. Both O and Ne abundances in bright PNe should be
useful probes for studying the chemical evolution of their host
galaxies.

Finally, it is necessary to understand what the mean oxygen
abundances in bright PNe represent. In the Magellanic Clouds,
bright PNe, H~{\sc ii} regions, and young stars all have similar
oxygen abundances \cite{richer93,hilletal95,venn99}. The same
situation is found in NGC 6822 \cite{vennetal01}.  The agreement
between stellar and nebular abundances implies that there are no
serious problems with the nebular abundance scale based upon
forbidden lines, since the stellar abundances are based upon
permitted lines. Bright PNe and H~{\sc ii} regions in NGC~3109,
Sextans A, and Sextans B likewise have similar oxygen abundances
\cite{magrinietal04,kniazevetal04}.  Therefore, in star-forming
systems, in general, bright PNe have oxygen abundances very
similar to those in the ISM.  Fewer systems without star formation
are available for similar study. The best case study is the bulge
of the Milky Way. There, bright PNe have a mean oxygen abundance
similar to the mean oxygen abundance in the stars
\cite{mcwilliamrich94,stasinskaetal98}. This result may hold more
generally for the entire PN population \cite{exteretal04}. In the
Fornax and Sagittarius dwarf spheroidals, the oxygen abundances of
their PNe are typical of the younger, more metal-rich stellar
populations.  In Sagittarius, this population appears to dominate
the mass \cite{cole01}.  In Fornax, the stellar population of
intermediate metallicity dominates the mass \cite{savianeetal00},
but Fornax's total stellar mass is so low that it is fortunate to
observe a single bright PN. Therefore, in systems without star
formation, the oxygen abundance in the bright PNe appears to be
representative of the mean oxygen abundance in the stars.

It is worthwhile trying to understand why this abundance dichotomy
arises in systems with and without star formation. Many processes
in stellar and galactic evolution can affect the population of
bright PNe. The means by which mass is lost by a PN precursor on
and immediately after the asymptotic giant branch (AGB) should
affect the eventual PN morphology. In the Milky Way, the
morphology of PNe correlates with their scale height above the
galactic plane \cite{garciaseguraetal02}, so morphology should
correlate with progenitor mass. Morphology could significantly
influence PN luminosity, since it will affect the angular
distribution of the nebular optical depth. There is a sensitive
interplay between the time scale to evolve from the AGB to the PN
phase and the nebular expansion velocity, since both affect the
density of the nebular shell and its consequent optical depth and
luminosity \cite{stasinskaetal98}.  A bright PN will only be
observed if the central star is hot and bright while the nebular
shell is sufficiently dense.  The time scale of the phase during
which the central star is bright is a strong function of the mass
of the central star, e.g., \cite{blocker95}, which, in turn, is a
function of the progenitor mass.  The time scale during which the
central star is bright also depends upon whether the it leaves the
AGB burning H or He, e.g., \cite{vassiliadiswood94}. Finally, the
time required for a star to evolve from the AGB to the PN phase
further depends upon the envelope mass remaining when the star
evolves off the AGB \cite{stanghellinirenzini00}, since this mass
must be disposed of through either a stellar wind or nuclear
burning.  All of these effects may well depend upon the progenitor
mass and metallicity. A key unknown is whether the evolution of
the nebular shell is correlated with that of the central star
\cite{villaveretal02,richeretal97}.

In a galactic context, chemical evolution usually ensures a
correlation between the masses and metallicities of the PN
progenitors.  This arises through the history of star formation,
which defines the mass range spanned by the PN progenitors and
their number distribution \cite{richeretal97}. In star-forming
galaxies, the PN progenitors span the entire range of masses
allowed.  Since the death rate in a stellar population is a strong
function of the population's age \cite{renzinibuzzoni86}, the
youngest populations can easily dominate the production of PNe, if
the star formation rate has been roughly constant (often a
reasonable approximation).  A high death rate alone is no
guarantee that the youngest stellar populations dominate the
production of bright PNe, as the highest-mass central stars evolve
so fast that it is statistically unlikely to observe such PNe
\cite{jacoby89,mendezsoffner97}.  In the LMC, the youngest stellar
populations apparently produce PNe fainter than do somewhat older
stellar populations \cite{stanghellini04}.  In DHSs, where star
formation ceased several $10^9$ years ago, the PN progenitors will
span a small range of masses, so the death rates among the
different stellar populations will be similar and each population
should contribute PNe in proportion to its mass, yielding mean
chemical abundances in the PN population very similar to those in
the stars. These arguments are in qualitative agreement with
attempts to model this process \cite{richeretal97} and illustrate
why the relations between abundances in bright PNe and other
tracers in galaxies with and without star formation might arise.

\section{The Chemical Evolution of DHSs}

A fundamental assumption in the analysis that follows is that the
bright PNe in all DHSs have mean oxygen abundances that are
representative of the mean oxygen abundance in their stars, as
occurs in the bulge of the Milky Way. The dwarf spheroidals Fornax
and Sagittarius are the only exceptions, where we assume that
their PN populations represent only their more metal-rich stellar
populations and the mean oxygen abundance for their entire stellar
content is reduced appropriately.  The results depend upon the
validity of this assumption.

\begin{figure}[t]
\begin{center}
\includegraphics[bb=85 270 735 720,width=12cm]{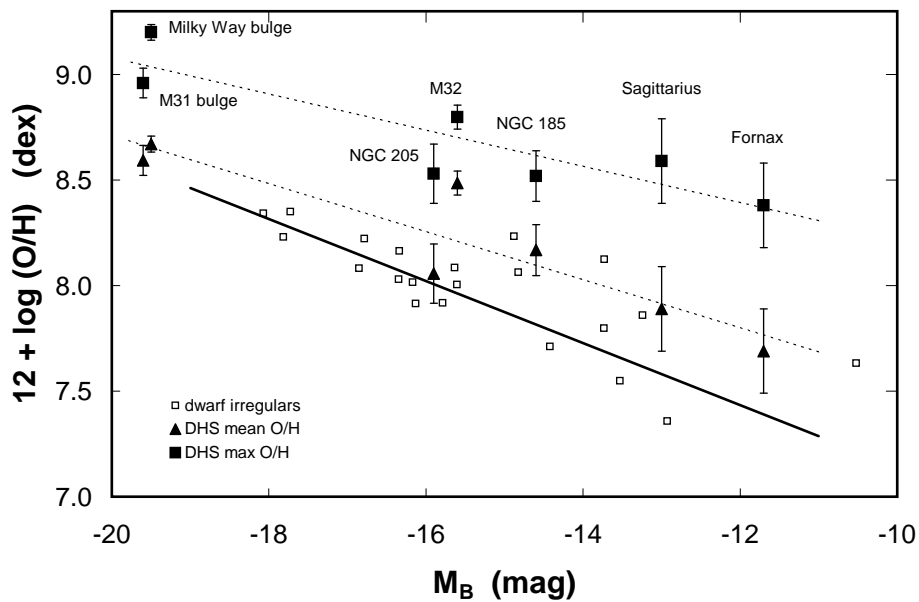}
\end{center}
\caption[The metallicity-luminosity relation for DHSs]{This figure
presents the relation between oxygen abundance and luminosity in
the $B$-band for a sample of DHSs.  The two dotted lines show the
mean and maximum oxygen abundances found for PNe in DHSs. For
reference, the same relation is shown for dwarf irregular
galaxies, which is based upon oxygen abundances measured in H~{\sc
ii} regions, and thus should be compared with the maximum
abundances observed in DHSs \cite{richermccall95}.}
\label{fig_mlrdhs}
\end{figure}

DHSs follow a well-defined relation between their oxygen
abundances and luminosities, as presented in Fig.
\ref{fig_mlrdhs}.  The relation is plotted using both the mean
oxygen abundance observed in the PNe in each DHS as well as using
the maximum abundance observed. The latter relation should
approximate the maximum oxygen abundances found in the ISM before
star formation ceased in these DHSs. These maximum abundances
should be approximately analogous to the oxygen abundances
observed in H~{\sc ii} regions in star-forming galaxies
\cite{richermccall95}.  In star-forming galaxies, this relation is
thought to arise because galaxies of different mass have converted
different fractions of their gas into stars, and so the remaining
ISM has been enriched more in those galaxies with the least gas
remaining \cite{leeetal03}. This same process is believed to have
occurred in DHSs, but that, at some point, the DHSs lost their
gas.

DHSs must have faded since they ceased forming stars.  As a
result, their abundance-luminosity relation has shifted to lower
luminosities over time.  It is easy to illustrate that fading a
dwarf irregular is insufficient of itself to reproduce the
abundance-luminosity relation followed by DHSs.  If a typical
dwarf irregular has a colour $(B-V)=0.5$\,mag, then after
$10^{10}$ years this galaxy will have faded by less than 2.5\,mag
\cite{girardietal00} if we approximate it as a single stellar
population with a metallicity $Z=0.008$, i.e., similar to the
metallicity of the LMC.  The approximation of a single stellar
population will exaggerate the amount of fading, since much of the
light is contributed by stellar populations with redder colours
that will fade less.  However, 5\,mag separate the
abundance-luminosity relation for dwarf irregulars and the maximum
abundance relation for DHSs.  Therefore, fading alone cannot
explain the shift of the abundance-luminosity relations for DHSs.

The abundance-luminosity relation for DHSs may also be offset if
their ISM was enriched more efficiently in newly synthesized
elements.  Preliminary attempts to model the mean oxygen
abundances and their dispersions find that it is difficult to
explain simultaneously the mean abundances and their dispersions
in DHSs unless DHSs were losing gas even while still forming stars
\cite{mccallricher03}. As a result, the enrichment of their ISM
proceeded at a faster rate than in today's dwarf irregulars
because of reduced dilution of the nucleosynthetic products
ejected by supernovae.  The effective yield of oxygen derived for
DHSs exceeds that derived for dwarf irregulars
\cite{mccallricher03,leeetal03}, and even that derived recently
for the inner parts of spiral galaxy disks
\cite{garnettetal04,bresolinetal04}, but well within the range
expected from stellar nucleosynthesis calculations
\cite{koppenarimoto91}.

\begin{figure}[t]
\begin{center}
\includegraphics[bb=90 270 740 720,width=12cm]{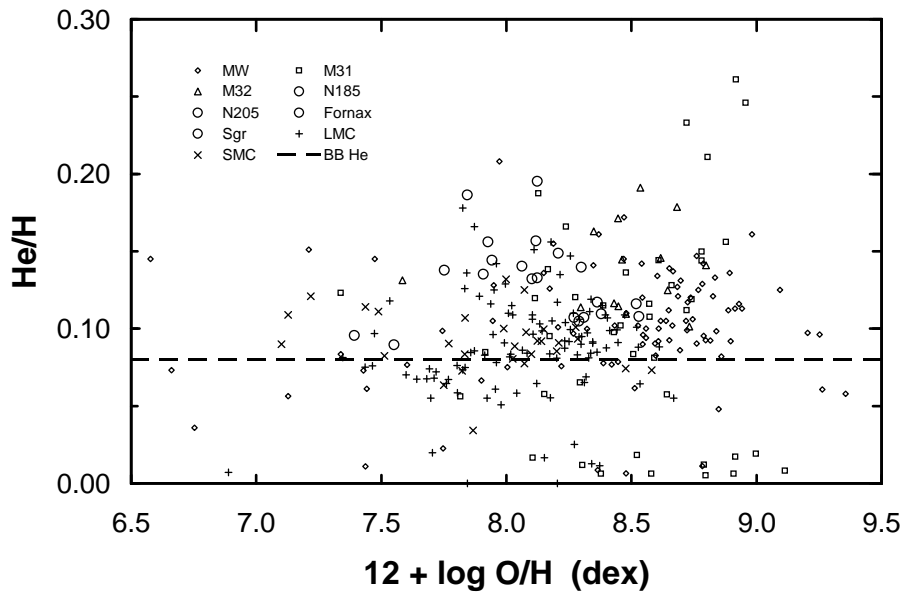}
\end{center}
\caption[]{Helium abundances are plotted as a function of the
oxygen abundance for the PNe in various galaxies.  At a given
oxygen abundance, the helium abundance is higher in the PNe in M32
and the dwarf spheroidals than it is in the PNe in the Magellanic
Clouds.  Given that mixing and nucleosynthesis have the same
effects upon the PN progenitors in all galaxies (Fig.
\ref{fig_neo}), processes related to the evolution of the host
galaxies, such as the enrichment of the ISM, should be responsible
for the higher helium abundances in the DHSs.  The very helium
abundances (near zero) are the result of missing ionization
stages.  The data are adopted from
\cite{stasinskaetal98,jacobyciardullo99,rothetal04}.}
\label{fig_heo}
\end{figure}

The helium abundances of the PNe in DHSs offer additional evidence
in favour of efficient enrichment of their ISM. In Fig.
\ref{fig_heo}, the helium abundances in PNe in different galaxies
are plotted as a function of their oxygen abundances.  At a given
oxygen abundance, the PNe in M32 and the dwarf spheroidals,
Fornax, Sagittarius, NGC 185 and NGC 205 have higher helium
abundances than the PNe in the Magellanic Clouds. This effect is
not so obvious for the bulges of the Milky Way or M31, but these
are also the DHSs that fall closest to the abundance-luminosity
relation for dwarf irregulars in Fig. \ref{fig_mlrdhs}. If the
similar scatter in N/O is taken as evidence that mixing and
nucleosynthesis in the PN progenitors was similar in all galaxies
(Fig. \ref{fig_neo}), the cause of the high helium abundances in
the DHSs should then be the result of the evolution of their host
galaxies, and not the PN progenitors.  Efficient enrichment of the
ISM is such a process.  In fact, gas outflow should affect helium
enrichment more than oxygen enrichment because the production of
helium is dominated by stars of lower mass (and which live longer)
than is the case for oxygen \cite{maeder92}, and so the ISM has
more time to evolve (lose mass) during the production of helium.

\begin{figure}[t]
\begin{center}
\includegraphics[bb=85 270 735 720,width=12cm]{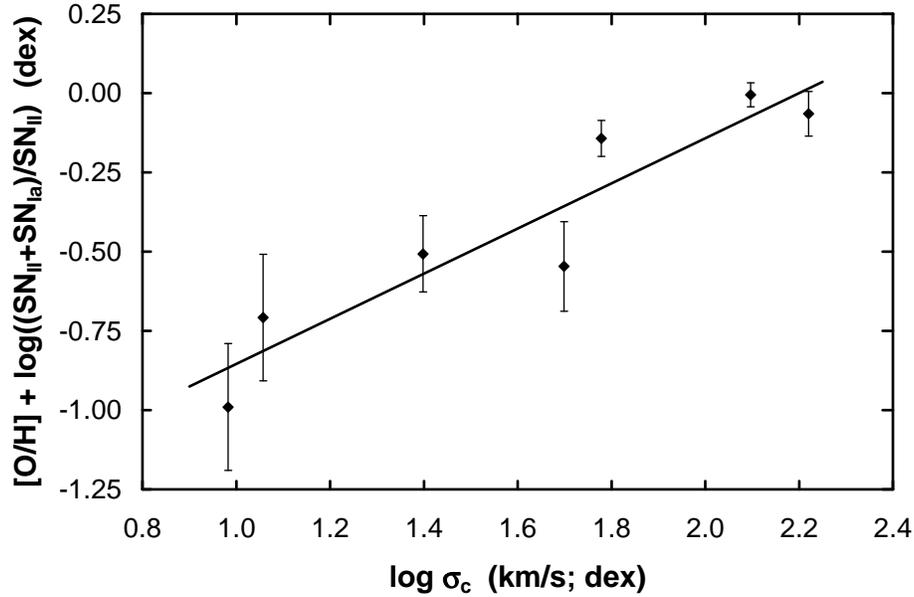}
\end{center}
\caption[]{Oxygen abundances are plotted versus velocity
dispersion in DHSs.  The oxygen abundances have been corrected to
account for the energy injection from type Ia SNe.  This relation
arises naturally if supernova-driven winds are the mechanism that
terminates chemical evolution in DHSs.} \label{fig_oxsig}
\end{figure}

Supernova-driven winds are commonly believed to have removed the
gas from DHSs \cite{larson74}.  The basis for this belief is the
correlation between metallicity indices and the velocity
dispersion, e.g., the Mg$_2-\sigma$ relation \cite{benderetal93}.
SNe are responsible for injecting newly-synthesized elements into
the ISM, but they likewise inject large quantities of energy into
the ISM.  If the energy injection rate exceeds the dissipation
rate, the thermal energy of the ISM increases and it may
eventually exceed its gravitational binding energy.  At that
point, the ISM begins to flow out of the galaxy in a wind,
presumably halting star formation at the same time.  Massive
galaxies are enriched more than low-mass galaxies since more SNe
are required to initiate a wind.  In this scenario, the chemical
enrichment of a galaxy is related to the velocity dispersion of
its stars only because both are controlled by the gravitational
potential. Since oxygen is produced by core collapse SNe, the
oxygen abundance is a faithful tracer of the energy they injected
into the ISM.  However, star formation in DHSs may have extended
over a period long enough that type Ia supernovae also contributed
energy and matter to the ISM.  One may correct the oxygen
abundances to account for the contribution of type Ia SNe using
the O/Fe ratio observed in stars \cite{richeretal98}, since type
Ia SNe contribute significantly to iron enrichment, but not to
oxygen enrichment. Fig. \ref{fig_oxsig} presents these corrected
oxygen abundances plotted as a function of the velocity dispersion
in their host galaxy.  A correlation is found that is consistent
with expectations if the chemical evolution of DHSs was terminated
by the action of supernova-driven winds.

The role of mergers in the formation of DHSs is difficult to
discern from these results. For gas-rich, major mergers
(components with similar masses), the chemical evolution is
changed very little compared to a monolithic formation scenario,
since a large fraction of the star formation occurs after
assembling the baryonic mass. Major mergers that are gas-poor
would flatten the abundance-luminosity relations in Fig.
\ref{fig_mlrdhs} and the abundance-velocity dispersion relation in
Fig. \ref{fig_oxsig}, for most of the stars would form in a
shallower gravitational potential than that in which they now
reside.  The slope and not the intercept of the relations is
affected because the smallest DHSs considered here are too small
to be the products of mergers themselves.  Minor mergers
(components with very different masses) would not be
distinguishable in Figs. \ref{fig_mlrdhs} and \ref{fig_oxsig}
regardless of their gas content.  Clearly, the only scenario ruled
out is the assembly of large DHSs entirely from a large number of
small, gas-poor systems.  This would produce a DHS that falls
below the relations in Figs. \ref{fig_mlrdhs} and \ref{fig_oxsig},
which has not been observed so far.  Basically, one would find the
same oxygen abundances in all DHSs, regardless of mass.

\section{Conclusions}

Two observations indicate that bright PNe in DHSs should be useful
probes to infer the chemical evolution of these galaxies. First,
the precursors of bright PNe do not significantly modify their
original oxygen abundances. Second, the mean oxygen abundance
measured in bright PNe appears to be representative of the average
oxygen abundance for the entire stellar population in DHSs.
Nonetheless, there are numerous uncertainties concerning the
production of populations of bright PNe and any advance in
quantifying the processes involved, at either the stellar or
galactic level, will improve the utility of PNe for studying the
chemical evolution of DHSs.

DHSs are found to follow a relation between oxygen abundance and
luminosity. This relation is shifted to higher abundances than
that followed by dwarf irregular galaxies, independently of
whether the mean or maximum oxygen abundances for the DHSs are
considered. Fading appears to be insufficient to explain the
shift.  DHSs appear to have incorporated their nucleosynthetic
production into subsequent stellar generations more efficiently
than did dwarf irregulars.

%

\end{document}